# On LoRaWAN Scalability: Empirical Evaluation of Susceptibility to Inter-Network Interference

Konstantin Mikhaylov, Juha Petäjäjärvi, Janne Janhunen
Centre for Wireless Communications
University of Oulu, P.O. BOX 4500, Oulu, Finland
Email: firstname.lastname@oulu.fi

*Abstract*—Appearing on the stage quite recently, the Low Power Wide Area Networks (LPWANs) are currently getting much of attention. In the current paper we study the susceptibility of one LPWAN technology, namely LoRaWAN, to the inter-network interferences. By means of excessive empirical measurements employing the certified commercial transceivers, we characterize the effect of modulation coding schemes (known for LoRaWAN as data rates (DRs)) of a transmitter and an interferer on probability of successful packet delivery while operating in EU 868 MHz band. We show that in reality the transmissions with different DRs in the same frequency channel can negatively affect each other and that the high DRs are influenced by interferences more severely than the low ones. Also, we show that the LoRa-modulated DRs are affected by the interferences much less than the FSK-modulated one. Importantly, the presented results provide insight into the network-level operation of the LoRa LPWAN technology in general, and its scalability potential in particular. The results can also be used as a reference for simulations and analyses or for defining the communication parameters for real-life applications.

## I. Introduction

The recent years were characterized by remarkable advances over the whole broad range of the wireless communication technologies, ranging from mobile broadband and up to a variety of niche and application-specific solutions, which have appeared on the stage quite recently. Among these newcomers the technologies, which can be collectively addressed as the Low Power Wide Area Networks (LPWANs), draw significant interest of Industry and Academy alike.

The landscape of the LPWAN technologies is quite diverse and is composed of the proper solutions (e.g., Sigfox, Ingenu/On-Ramp, Starfish, Cynet, Accellus, Telensa), open and semi-open technologies (e.g., LoRaWAN, Weightless), and few standards (IEEE 802.15.4k, LTE-M and NB-IoT recently adopted in Rel. 13 document). Albeit these technologies differ in respect to the technical solutions implied, their targets have much in common. Namely, they attempt to come up with the cost-effective means for wireless communication for massive deployments of autonomous machines. Note, that the cost efficiency here has several notations. First, it implies the low cost of the transceiver itself thus calling to make it as simple as possible. Second, this urges to reduce the operational costs of a device, e.g., by maximizing its lifetime. Third, the low installation and exploitation cost of the infrastructure, facilitating the use of long-range communication resulting in low density of infrastructure, is also of extreme importance. This makes the LPWANs the perfect solution for the variety of the infrastructure and real-estate monitoring applications. To give a practical example, large real estates have numerous locations where wireless sensing may be employed like monitoring for the leakages of water pipes or air quality control. Given potentially large number of devices operating in the license free ISM bands, co-channel interference might become an issue at some point. Therefore, in this work we study the robustness of one LPWAN technology, namely LoRaWAN, to inter-network interference.

In the academic community, the interest towards the LPWANs has arisen only recently. One of the substantial difficulties when dealing with the LPWANs is the lack of the information in the open access, especially when this comes to the proprietary radio technologies. Even though, in [1]-[3] the authors reviewed the available LPWAN technologies. Meanwhile, thanks to the availability of the protocol specification and the mature deployment state, the Long Range (LoRa) Wide Area Network (WAN) - LoRaWAN technology – is probably the most well studied LPWAN technology as of today. The empirical validation of the LoRa technology coverage has been reported e.g., in [4]-[6]. Some characteristics of the dynamic behavior and energy consumption were reported in [7]. Finally, most relevant to this work, the problem of the LoRaWAN scalability has been approached in [8]-[12].

In [8] were analyzed the LoRaWAN protocol and assessed the fundamental throughout limits, as well as were evaluated the number of devices which can be supported in a single LoRaWAN cell for few illustrative applications (assuming orthogonality of the different data rates (DRs)). In [10] Georgiou and Raza employ the tools of stochastic geometry to analyze the performance of a LoRa network under inter-network interferences, showing that the performance decays exponentially with increase of the number of devices in the network. The two key assumptions in this paper were the possibility of receiving the colliding packets with equal spreading factors (SFs) if the desired signal is 6 dB stronger, and the orthogonality of signals with different SFs. The assumption of the different SFs being orthogonal and not affecting each other has been also employed by the authors in [12], in which the two mechanisms for mitigating the inter-network LoRa interferences were studied by means of simulations. The conclusions driven are that the interferences can substantially degrade the LoRa network performance and that the deployment of multiple base stations can help mitigating this problem more efficiently than the use of directional antennae. In [9] the simulations were employed to analyze the robustness of the chirp spread spectrum (CSS) and ultra-narrow band (UNB) LPWAN technologies subject to interferences. Importantly, not only the case of in-network, but also of external and mutual interferences was explored. Finally, in [11] the authors first experimentally study the operation of LoRa under interferences, showing the presence of the capture effect enabling reception of the packets under interfering signal having the same SF (similarly to the other works, the authors expect that there is no interference from the signals with different SF). Then these results are used in a simulation to assess the number of the nodes which can be handled by one, and by multiple gateways.

As can be seen from the previous discussion, most of the current works dealing with the LoRaWAN scalability use analytical approach or simulations. The results of the real-life experiments reported in [11] are limited to the measurement of sensitivity and illustration of the capture effect. Therefore, in order to breach this gap, in this paper we report results of an extensive empirical experimentation campaign intended to shed some light on the real-life performance of LoRaWAN devices under the in-network interferences. Namely, we aim at validating the orthogonality of the transmissions using the different SFs, as well as to check the effect of mutual interference between the LoRa and GFSK modulated signals.

The paper is organized as follows. In Section II we briefly overview the LoRaWAN technology. In Section III we first detail our experimental setup, present the obtained results and discuss some of them. Finally, Section IV concludes the paper and highlights the major findings.

## II. LoRaWAN Technology in Brief

The LoRaWAN solution consists of the two major components. The first one is the LoRaWAN network protocol, defined in the LoRaWAN specification [12]. The network is deployed in a star-of-stars topology, where the end devices (EDs) exchange their data to the gateways (GW), which relay them to a central network server (NS) via a backhaul IP-based connection. The communication between the EDs and a GW is done wirelessly, using one of the several modulation-coding schemes, called DRs, the set of which depends on the operation location and the local frequency regulations. E.g., in European 868 MHz band this set includes 8 options listed in Table I. A DR used by a device may be changed, thus enabling to trade the on-air time for the communication range. Multiple frequency channels should be enabled in each LPWAN setup and used by the EDs uniformly. The specification defines also three types of the EDs, labeled A, B and C. The former ones access the radio channel in Aloha fashion, selecting one of the available channels randomly. The two receive windows (RW) are opened after each uplink message to enable for downlink. The rest of the time the radio can be powered down. The type B devices in addition to these to RWs, have pre-scheduled RWs. Finally, class C devices do not use low power modes and stay in receive when they do not transmit. Since not employing any listen before talk technique, the LoRaWAN devices have to comply the duty cycle restrictions imposed by the frequency regulatory authorities. Due to this, each ED tracks its on-air time and backs off the transmission accordingly to meet the regulations. The adaptive DR (ADR) feature can be used to enable NS control the power and the DR used by a particular ED, if enabled.

The second key component of the LoRaWAN is the LoRa modulation scheme. In essence, the LoRa modulation is a CSS-based scheme with chirp signal constantly varying in frequency [1]. The raw physical rate is thus defined as [1]: $Rb=SF\ BW/2^{SF}$ where $SF$ is the spreading factor used and $BW$ is the bandwidth (refer to Table I). Additionally, LoRa employs error correcting codes (typically with the fixed rate of 4/5 [14]).

TABLE I. LoRaWAN DRs for EU 868 band [14]

| DR | Modulation | LoRa SF | Bandwidth | PHY rate, bps | Max MAC payload, bytes |
|---|---|---|---|---|---|
| 0 | LoRa | 12 | 125 | 250 | 59 |
| 1 | LoRa | 11 | 125 | 440 | 59 |
| 2 | LoRa | 10 | 125 | 980 | 59 |
| 3 | LoRa | 9 | 125 | 1760 | 123 |
| 4 | LoRa | 8 | 125 | 3125 | 230 |
| 5 | LoRa | 7 | 125 | 5470 | 230 |
| 6 | LoRa | 7 | 250 | 11000 | 230 |
| 7 | GFSK | 50 000 bit/s rate | | 50000 | 230 |

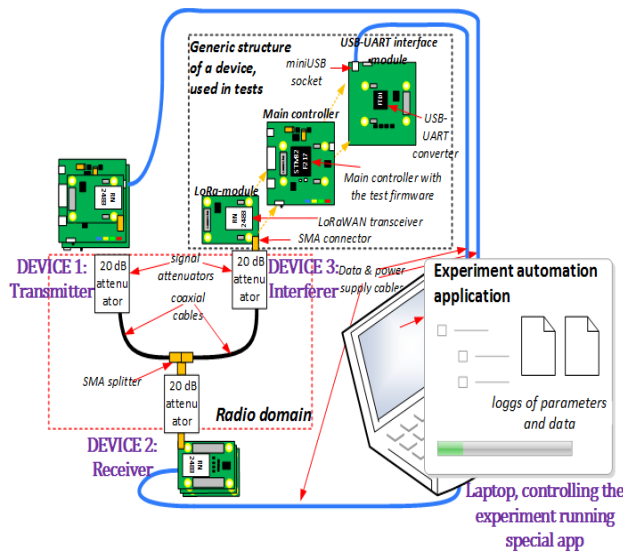

(a) structural diagram

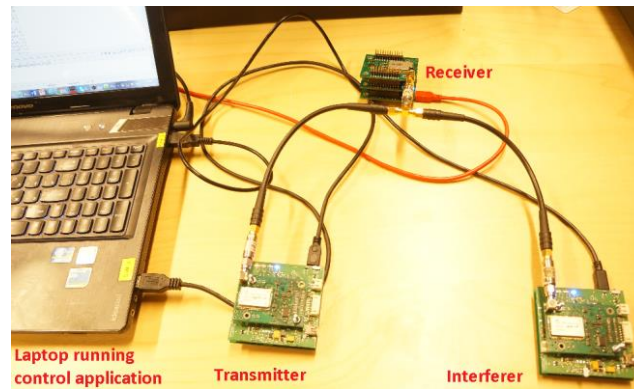

(b) Photo of the test bed
Fig. 1. Experimental setup

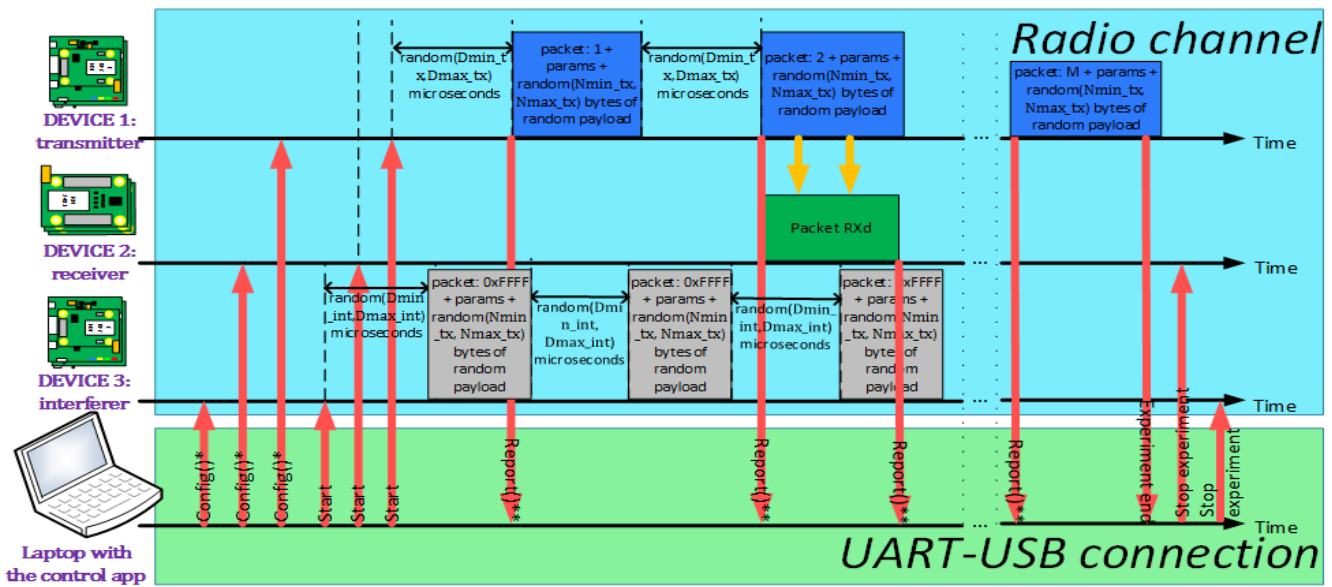

Fig. 2. Sequence of commands and messages while testing a single mode.

\*- parameters of a configuration command: role, DR, Ptx[transmit power], interpacket delay (Dmin and Dmax), packet payload (Nmin and Nmax), number of packets to send (M)
\*\* parameters of a report: full packet payload of received/transmitted packet; for received packet – also the estimated link quality

## III. EXPERIMENTAL EVALUATION OF LoRaWAN SUSCEPTIBILITY TO INTERFERENCES

### A. Experimental setup

The tests were conducted using the modular Wireless Sensor and Actuator Network (WSAN)/IoT platform developed at the Centre for Wireless Communications (CWC) of the University of Oulu ([15],[16]). Specifically for these tests, three modules based on the RN2483 LoRaWAN transceivers[1] from Microchip were designed (the RN2483s were programmed with firmware v1.0) and built. One of the features of these transceivers, which was employed in the test, is the possibility to disable the LoRaWAN MAC stack and to enable the direct control over the radio from the controller [18]. The embedded software implementing the low level drivers as well as the test applications were also developed, along with the Java based experiment automation application.

The setup of our tests is detailed in Fig. 1 and the experiment composition for a single tested mode is depicted in Fig. 2. Three devices were used in the experiments, namely the transmitter, the receiver and the interferer. The antenna inputs/outputs of the transceiver modules were connected through coaxial cables. The 20 dB attenuators were installed at the antenna input/output of each transceiver. The major reasons for not conducting the experiments in over-the-air channel were: 1) avoidance of interferences from external systems and 2) escaping the restrictions regarding the duty cycle imposed by the regulations. All three devices were instrumented with USB-UART modules and connected to USB ports of a laptop running the Java test control program. The USB interfaces were used to control the experiments and collect the data from the devices.

TABLE II. EXPERIMENT PARAMETERS

| DR | TX power[1], dBm | Transmitter | | Interferer | |
|---|---|---|---|---|---|
| | | Payload, bytes | Interpacket delays, ms | Payload, bytes | Interpacket delays, ms |
| 0 | 0,3,6,9,14 | 10..51 | 65.5-69.6 | 10..51 | 10-20 |
| 1 | 0,3,6,9,14 | 10..51 | 65.5-69.6 | 10..51 | 10-20 |
| 2 | 0,3,6,9,14 | 10..51 | 65.5-69.6 | 10..51 | 10-20 |
| 3 | 0,3,6,9,14 | 10..115 | 110.6-114.7 | 10..115 | 10-20 |
| 4 | 0,3,6,9,14 | 10..127 | 122.9-127.0 | 10..242 | 10-20 |
| 5 | 0,3,6,9,14 | 10..127 | 122.9-127.0 | 10..242 | 10-20 |
| 6 | 0,3,6,9,14 | 10..127 | 122.9-127.0 | 10..242 | 10-20 |
| 7 | 0,3,6,9,14 | 10..63 | 65.5-69.6 | 10..63 | 10-20 |

[1] – valid both for transmitter and interferer

The experiments were composed of two phases. In the first one, only the transmitter and the receiver were used. The transmitter was configured to send packets using various transmit powers and modulation parameters. The receiver, operating using the same settings as the transmitter, attempted to receive these packets. Each of the packets consisted of the unique sequence number, payload length and the time gap between the current and the previous packet, and a randomly generated payload. The length of the payload for each packet was also generated randomly in between the minimum and maximum values listed in Table II. The transmitted and received packets (the latter included also the link quality estimation) were reported by the devices to the Java application and logged in the file system of the laptop for further analysis. Note that the delays between sequential packets during the experiments were *random* and included two components. First – a delay due to serial communication over UART and processing. And second – the artificial additional random delay generated with the granularity of 1 µs within the ranges specified in Table II.

---

[1] according to [16] RN2483 is the world's first transceiver to pass the LoRaWAN™ Certification Program

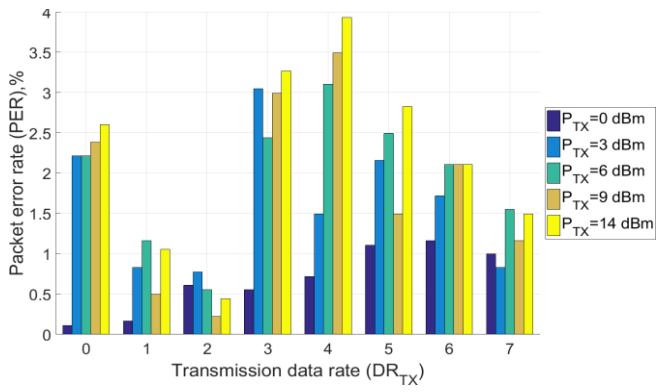 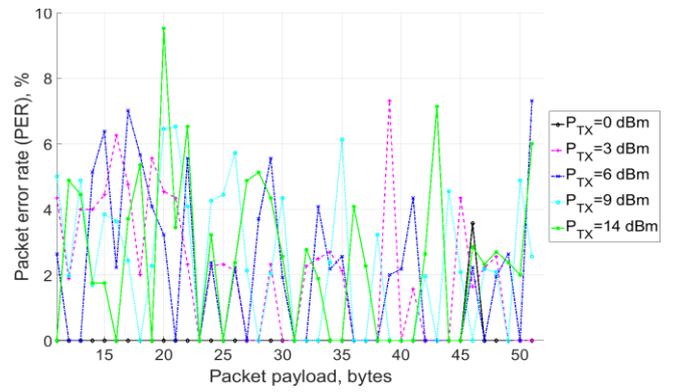

a) Average PER

b) Effect of payload on PER (DR0)

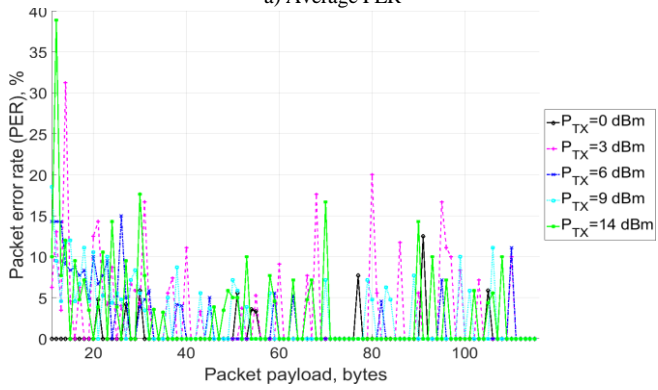 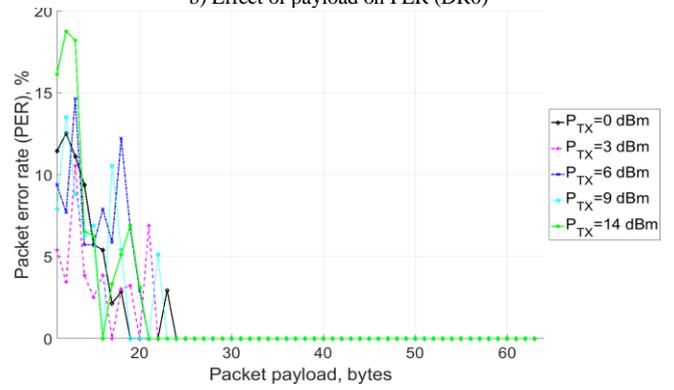

c) Effect of payload on PER (DR3)

d) Effect of payload on PER (DR7)

Fig. 3. Results of the phase one measurements (only transmitter and receiver are active) (color high resolution figure online)

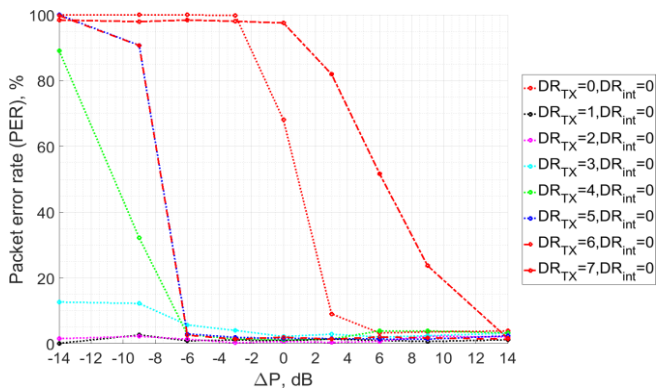 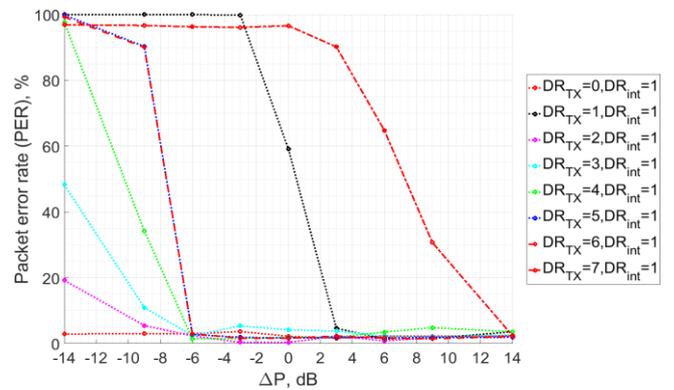

a) Effect of signal-interference power on PER for interferer with DR0

b) Effect of signal-interference power on PER for interferer with DR1

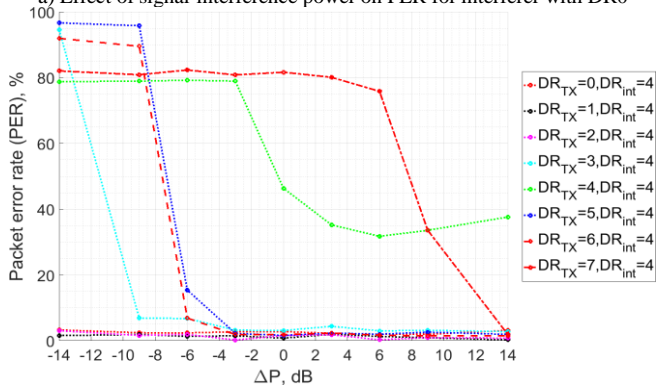 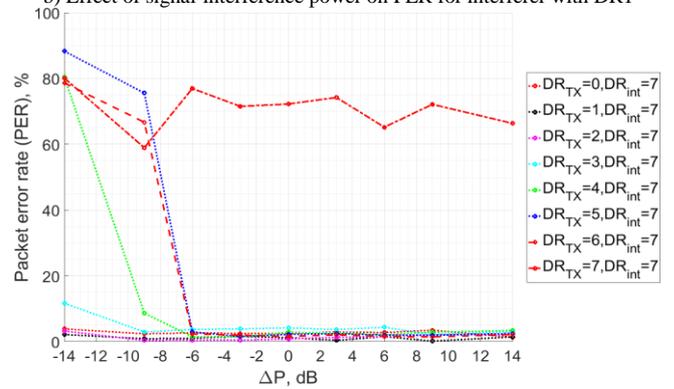

c) Effect of signal-interference power on PER for interferer with DR4

d) Effect of signal-interference power on PER for interferer with DR7

Fig. 4. Selected results of the phase two measurements (transmitter, interferer and receiver are active) (color high resolution figure online)

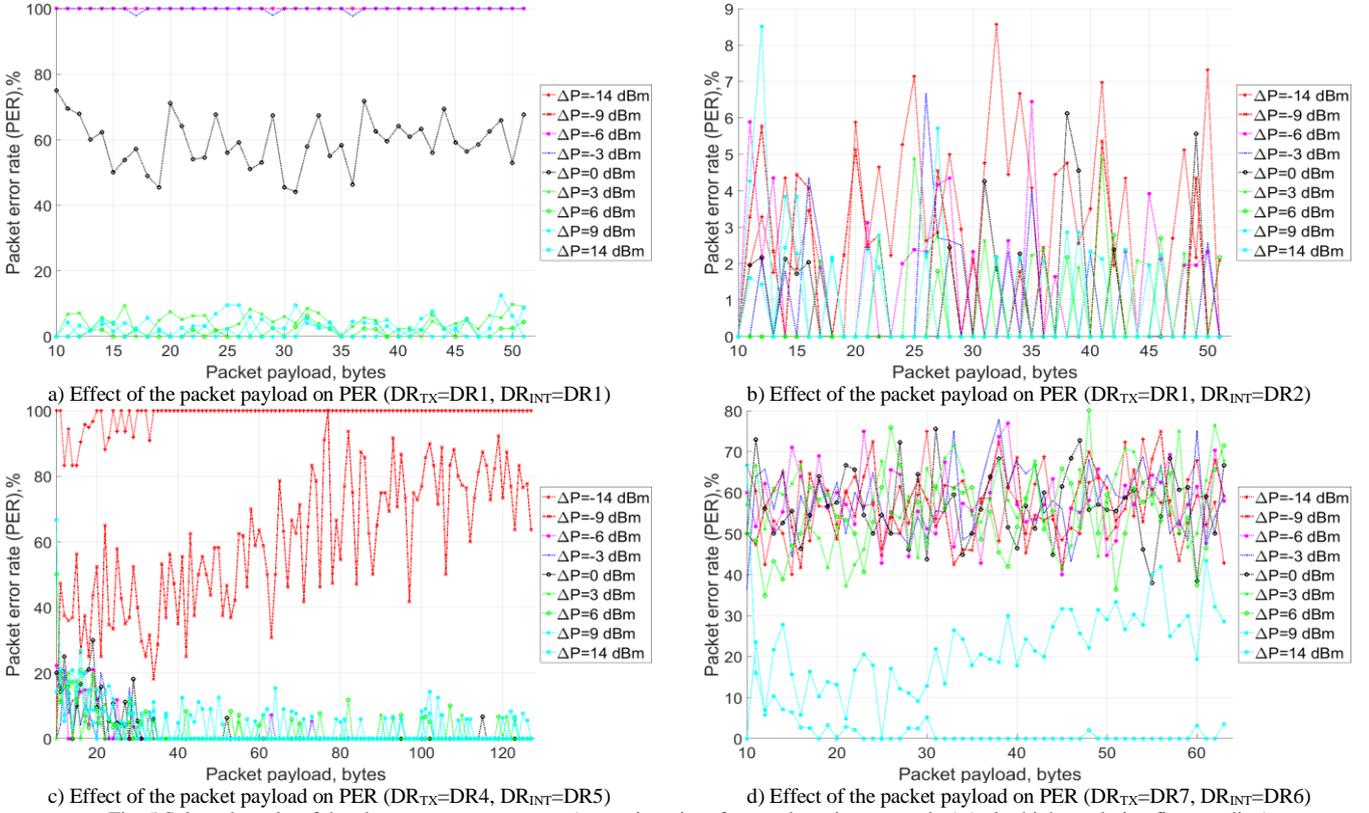

a) Effect of the packet payload on PER ($DR_{TX}$=DR1, $DR_{INT}$=DR1)
b) Effect of the packet payload on PER ($DR_{TX}$=DR1, $DR_{INT}$=DR2)
c) Effect of the packet payload on PER ($DR_{TX}$=DR4, $DR_{INT}$=DR5)
d) Effect of the packet payload on PER ($DR_{TX}$=DR7, $DR_{INT}$=DR6)
Fig. 5 Selected results of the phase two measurements (transmitter, interferer and receiver are active) (color high resolution figure online)

During the second phase, in addition to the transmitter and the receiver the interferer was employed. The operation of interferer was similar to the one of transmitter subject to minor differences in the parameters, as detailed in Table II.

In total, 40 (i.e., 8 DRs * 5 power settings) modes in the first phase and 576 (i.e., 8 DRs for transceiver * 8 DRs for interferer * 9 different interferer/transmitter power level combinations) experimental modes in the second phase were checked. For each of these modes 2000 packets were transmitted by the transmitter (the number of packets sent by interferer was not restricted). In total the experiments lasted for over 100 hours non-stop and involved transmission of over 2,5 million radio packets.

*B. Discussion of the Results*

Due to the space restrictions, only a small and the most illustrative portion of all the results are presented. The selected results of the first phase are depicted in Fig. 3. As can be seen from Fig. 3(a) for each of the tested modes some packets were lost. Nonetheless, the number of lost packets never exceeded 4% of the total. Interestingly, the higher transmit power often resulted in more packets being lost. This may have been the effect of power saturation, but further experiments are needed to evaluate this hypothesis. Another interesting observation can be derived from Figs. 3(b)-3(d) depicting the effect of the payload size on the packet error rate (PER). As can be seen, for low DRs (namely DR0-DR2) the packets are lost rather uniformly. Starting from DR3, the packets with low payloads get lost more often and for DR7 all the packets lost had small payload. Unfortunately, the lack of information about the internal design of the used transceiver does not enable us to make justified conclusions about the reasons causing this.

The selected results of the second phase are illustrated in Figs. 4 and 5. The former figure reveals the effect of the power difference between the target and the interfering signals on the PER. The latter illustrates the effect of the payload on the PER for the different power levels and DR settings. As one can clearly see from Figs. 4(a)-(d) the transmissions with different DRs in the same channel affect one another. For example, as can be seen from Fig. 4(a) the interferer operating with DR0 reduced the probability of successful packet delivery for devices operating with DR2-DR6 in the case if the power of the interfering signal exceeded the one of the target device by more than 6 dB at the receiver. Also one can see that the transmission with higher DRs are affected stronger than the ones with low DRs (compare, e.g. DR4 and DR3). In the case if both a target device and an interferer operating with DR0, about 1/3 of the packets were delivered successfully for equal signal power levels, and more than 90% when the target signal was 3 dB stronger than the one of the interferer. The analysis of the other subfigures shows that interferer typically negatively affects the communication of the devices operating with DRs higher than the DR of the interferer and may affect the communication for 1-2 lower DRs (e.g., consider DR3 in Fig. 4(c)). Also the obtained results show that in case if the interferer operates with the same DR as the target device, a higher power gap is needed for higher DRs to correctly receive a packet. Note also that with high DRs some packets are delivered even when the power of the interferer exceeds one of the target. Most likely this is the effect of the packet duration – the relatively short packets get successfully delivered in the time gaps between the frames of the interferer. The presented results clearly show the difference between the DRs based on GFSK and LoRa modulation. The former ones are negatively affected by any LoRa interference and require the target signal to be 9 dB stronger than the interfering one to get more than 2/3 of the packets through. The LoRa-modulated signal (see Fig. 4(d)) is much less affected by the GFSK interference; if the interfering GFSK signal is less than 6 dB stronger than the LoRa signal – it has no visible effect.

Analyzing the effect of the packet payload on the PER under interference, once can see that there are few different trends. In case if the interferer has the DR matching the one of the target signal – the length of the payload does not affect the PER. The same happens when the target signal is transmitted at low DR (DR0-DR2, see Fig. 5(b)). For the higher target signal DR settings (refer to Figs. 5(c) and 5(d)) the trends vary. For some power levels (e.g., consider ΔP=-9 dB and -14 dB in Fig. 5(c)) the packets with higher payload are less likely to be delivered correctly. Meanwhile, when the signal from the interferer is weaker, the short packets are lost more often (see, e.g., ΔP=-6 dB to +14 dB in Fig. 5(c)).

## IV. Conclusions

To the best of authors' knowledge, the current paper is the first attempt to empirically investigate the effect of inter-network interference on performance of the recently developed LoRa LPWAN technology. By the means of excessive measurements executed with certified commercial LoRaWAN transceivers, we characterized the effect of the modulation coding schemes (i.e., DRs) of the transmitter and the interferer on probability of successful packet delivery. Based on our results the following conclusions were derived. 1) In contrast to the common analytical assumption, in real life the LoRaWAN communications with the different DRs in the same frequency channel do negatively affect each other. 2) The packets sent with high DRs are affected by the interferences stronger than the ones sent with low DRs. 3) If the interferer operates with the same DR as the target device – there is still a probability of receiving the correct packet, which depends both on the difference in the power levels of the two signals and the used DR. 4) If LoRa modulation is used, and the interfering signal is encoded at different DR and is less than 6 dB stronger than the target signal – there are good chances (>80%) to receive the target signal. 5) The reception of a LoRa-modulated signal can be affected by a GFSK interferer if the latter is more than 6 dB stronger than the LoRa signal. The reception of a GFSK signal under LoRa interference is likely if the GFSK signal is much stronger (>9 dB) than the interfering LoRa signal. 6) The payload size can affect the probability of packet reception in various ways, depending on the DRs and the power levels of the interfering and the target signals.

The obtained results are valuable from a number of perspectives. First, they provide insight into the network-level operation of the LoRa LPWAN technology in general, and its scalability potential in particular. In this respect, the presented results enable a more accurate analysis of the number of devices in an LPWAN or of the quality of service provided by the system. Second, the presented results show the importance and provide valuable background information for designing an adaptive data rate (ADR) and transmit power control mechanism for LoRaWAN. Third, the presented results can enable more reasonable selection of the communication parameters to be used for a particular application.

In future, we consider extending the measurements by including more nodes. Also we plan to investigate further the strange effect of the transmit power on the packet error rate.

ACKNOWLEDGMENT

This work has been partially funded by the Finnish Funding Agency for Innovation (Tekes) through VIRPA-C project.